\documentclass[aps,pre,twocolumn,floatfix,showpacs,groupedaddress]{revtex4-1}
\usepackage{amssymb}
\usepackage{amsmath}
\usepackage{bm}
\usepackage{graphicx}
\usepackage{url}
\usepackage{xcolor}
\DeclareMathOperator{\Tr}{Tr}
\DeclareMathOperator{\diag}{diag}

\begin{document}


\title{Computational molecular field theory for nematic liquid crystals}


\author{Cody D. Schimming}
\email[]{schim111@umn.edu}
\affiliation{School of Physics and Astronomy, University of Minnesota, Minneapolis, Minnesota 55455, USA}

\author{Jorge Vi\~nals}
\affiliation{School of Physics and Astronomy, University of Minnesota, Minneapolis, Minnesota 55455, USA}

\date{\today}

\begin{abstract}
Nematic liquid crystals exhibit configurations in which the underlying ordering changes markedly on macroscopic length scales. Such structures include topological defects in the nematic phase and tactoids within nematic-isotropic coexistence. We discuss a computational study of inhomogeneous configurations that is based on a field theory extension of the Maier-Saupe molecular model of a uniaxial, nematic liquid crystal. A tensor order parameter is defined as the second moment of an orientational probability distribution, leading to a free energy that is not convex within the isotropic-nematic coexistence region, and that goes to infinity if the eigenvalues of the order parameter become non-physical. Computations of the spatial profile of the order parameter are presented for an isotropic-nematic interface in one dimension, a tactoid in two dimensions, and a nematic disclination in two dimensions. We compare our results to those given by the Landau de-Gennes free energy for the same configurations and discuss the advantages of such a model over the latter.
\end{abstract}


\maketitle

\section{\label{Introduction}Introduction}
Liquid crystals represent an interesting opportunity to study a unique interplay between topology, anisotropy, and elasticity in materials. The entropy driven local ordering of rod-like molecules accounts for anisotropic optical and transport properties even in homogeneous nematics. Furthermore, external fields or topological defects can distort the local ordering of the molecules giving rise to several elastic modes \cite{deGennes75,selinger19}. The ability to quantitatively model these complex features of liquid crystals is imperative to address recent applications, including electrokinetics of colloidal particles or biological materials \cite{lazo14,peng15,peng18}, surface and texture generation and actuation in nematic surfaces \cite{most15,baba18}, systems of living nematics \cite{genkin17}, and stabilization of liquid shells \cite{hokmabad19}. 

Liquid crystals generally belong to one of two main classes: Thermotropics are short molecules that undergo ordering through changes in temperature, while lyotropics are more complex molecules or assemblies of molecules in solvent that order through changes in concentration. Thermotropics have been extensively studied, both theoretically and experimentally, due to their applications in displays \cite{deGennes75,yeh09}. However, because of their small characteristic length scale, the fine structure of defects and two phase domains (commonly referred to as tactoids) are generally beyond the resolution of standard optical techniques. On the other hand experimental studies of defect core structures and tactoids have been recently undertaken in so called lyotropic chromonic liquid crystals. These materials are composed of disc-like molecules that stack to form rod-like structures \cite{collings10,collings15}. The characteristic length scale that determines the size of defects and tactoid interfacial thickness in chromonics are thousands of times larger than those in thermotropics, and hence are readily observable with conventional optical techniques. Such experiments have revealed anisotropic geometries of the order parameter near the core of defects, and \lq\lq cusp-like'' features on the interface of tactoids \cite{kim13,zhou17}. 

To mathematically model a liquid crystal in its nematic phase a unit vector \(\mathbf{n}\), the director, is typically defined to characterize the local orientation of the molecules. Because the molecules are apolar, any model involving \(\mathbf{n}\) must be symmetric with respect to \(\mathbf{n} \to -\mathbf{n}\). Distorted nematic configurations are described by three independent elastic modes: splay, twist, and bend. The energy cost of each mode is associated with three elastic constants \(K_1\), \(K_2\), and \(K_3\) in the Oseen-Frank free energy \cite{selinger19,frank58}. Models and computations often assume that these constants are equal, though it has been shown for chromonics that the values of all three constants are widely different for the relevant range of temperatures and molecular concentrations \cite{zhou14}. Additionally, topological defects and tactoids lead to large distortions of the underlying order. To model defected configurations using the Oseen-Frank free energy either a short distance cutoff is introduced,  and the defect core treated separately, or a new variable representing the degree of order of the molecules is  added to the free energy \cite{leslie66,ericksen91}. This new variable also has the effect of regularizing singularities at the core of defects. The method has recently allowed the study of tactoids within the coexistence region \cite{zhang18}.

Resolving the degree of orientational order and the orientation poses several challenges computationally, however. The director is undefined both at the core of defects and in the isotropic phase, and half-integer disclinations (the stable line defects in liquid crystals) cannot be adequately described computationally with a polar vector. Therefore, the model that is widely used to describe either disclinations or tactoids is the phenomenological Landau-de Gennes (LdG) free energy \cite{meiboom82,golovaty19,popanita97}. In the LdG framework, the order parameter is defined to be a traceless and symmetric tensor, \(\mathbf{Q}\), typically proportional to a macroscopic quantity, e.g. the magnetic susceptibility \cite{gramsbergen86,lubensky70}. The free energy is then assumed to be an analytic function in powers of \(\mathbf{Q}\). To model spatial inhomogeneity, an expansion in gradients of \(\mathbf{Q}\) is typically added to the free energy. Such an expansion in gradients can be mapped to the elastic modes in the director \(\mathbf{n}\) in the Oseen-Frank elastic energy \cite{selinger19}. 

The validity of the LdG free energy in regions of large variation of the order is not well understood, and it has been shown that the simplest LdG elastic expansions that capture differences in the Oseen-Frank constants result in unbounded free energies \cite{longa87,ball10}. Therefore, when working in the LdG framework, one must introduce more computationally complex assumptions to bound the free energy. In this work, we present an alternative field theoretic model of a nematic liquid crystal that is based on a microscopic description, and that allows for anisotropic elastic energy functionals that can capture the elasticity observed in chromonics. The model presented here is a computational implementation of the model introduced by Ball and Majumdar \cite{ball10}, which itself is a continuum extension of the well known Maier-Saupe model for the nematic-isotropic phase transition \cite{maier59}. The Maier-Saupe model is a mean field molecular theory in which the orientation of the molecules of the liquid crystal is described by a probability distribution function, so that each molecule interacts only with the average of its neighbors. Below, we define \(\mathbf{Q}\) microscopically, based on a probability distribution that is allowed to vary spatially (as in the hypothesis of local equilibrium in nonequilibrium thermodynamics). Our ultimate goal is to develop a computationally viable implementation of the model for fully anisotropic systems. We present below the results of several proof of concept computations on various prototypical liquid crystal configurations, albeit in the one elastic constant approximation. All our results are compared with those from the LdG free energy for analogous configurations.

In Section \ref{Model} we briefly summarize the model as put forth in Ref. \cite{ball10} with minor adjustments to notation and conceptual understanding. In Section \ref{compMeth} we present the computational implementation of the model and derive the equations that are solved numerically. We also briefly discuss the conventions used to compare to the LdG free energy. In Section \ref{results} we compare the free energies of the model presented here with that given by LdG and show that they are both non-convex. We then present computational results from the model for a one dimensional nematic-isotropic interface, a two-dimensional tactoid, and a two-dimensional disclination. All of these are compared to results given by LdG. Finally, in Section \ref{Conclusion} we summarize and discuss the computational model and results, and discuss future potential for the model.  

\section{\label{Model}Model}
Following Ref. \cite{ball10}, we consider a tensor order parameter defined over a small volume at \(\mathbf{r}\)
\begin{equation}\label{MicroQ}
\mathbf{Q}(\mathbf{r}) = \int_{S^2} \big(\bm{\xi} \otimes \bm{\xi} - \frac{1}{3}\mathbf{I}\big) p(\bm{\xi};\mathbf{r}) \, d \bm{\xi}
\end{equation}
where \(\bm{\xi}\) is a unit vector in \(S^2\), \(\mathbf{I}\) is the identity tensor, and \(p(\bm{\xi};\mathbf{r})\) is the canonical probability distribution of molecular orientation in local equilibrium at some temperature $T$ at \(\mathbf{r}\). Due to the symmetry of the molecules, \(p(\bm{\xi};\mathbf{r})\) must have a vanishing first moment; hence, \(\mathbf{Q}\) is defined as the second moment of the orientational probability distribution. With this definition, the order parameter is symmetric, traceless, and, most importantly, has eigenvalues that are constrained to lie in the range \(-1/3 \leq q \leq 2 / 3\). The situation where \(q = -1/3, \, 2/3\) represents perfect ordering of the molecules (i.e. the variance of the distribution goes to zero), and is therefore interpreted as unphysical. We note that Eq. \eqref{MicroQ} can be generalized to biaxial molecules, that is, molecules that are microscopically plate-like, by appropriately changing the domain of the probability distribution to three Euler angles, and considering the second moment of the extended probability distribution. Such a description may be useful in studying similar defects and domains for biaxial molecules, as in Ref. \cite{chiccoli19}.

A mean field free energy functional of \(\mathbf{Q}(\mathbf{r})\) is defined by
\begin{equation} \label{FreeE}
F[\mathbf{Q}(\mathbf{r})] = H[\mathbf{Q}(\mathbf{r})] - T \Delta S
\end{equation}
where \(H\) is the energy of a configuration, and \(\Delta S\) its entropy relative to the uniform distribution. The energy is chosen to be
\begin{equation} \label{H}
H[\mathbf{Q}(\mathbf{r})] = \int_\Omega \Big(-\alpha\Tr[\mathbf{Q}^2] + f_e(\mathbf{Q},\nabla \mathbf{Q})\Big) \, d\mathbf{r}
\end{equation}
where \(\alpha\) is an interaction parameter, and \(f_e\) is an elastic energy. The term \(-\alpha \Tr[\mathbf{Q}^2]\) originates from the Maier-Saupe model, and incorporates an effective contact interaction that promotes alignment \cite{maier59,selinger16}. In the spatially homogeneous case \(f_e = 0\). The entropy is the usual Gibbs entropy
\begin{equation} \label{deltaS}
\Delta S = -  n k_B \int_\Omega \bigg(\int_{S^2} p(\bm{\xi};\mathbf{r}) \ln\Big(4 \pi p(\bm{\xi};\mathbf{r})\Big)\, d\bm{\xi}\bigg) \,d\mathbf{r}
\end{equation}
where \(n\) is the number density of molecules. It should be noted that the outer integral is on the physical domain of the system, and the inner integral is on the unit sphere, the domain of the probability distribution. This model, with these definitions, is equivalent to the Maier-Saupe model in the spatially homogeneous case \cite{maier59}. We extend the Maier-Saupe treatment to spatially nonuniform configurations by minimization of Eq. \eqref{FreeE} subject to boundary conditions that lead to topological defects in the domain, or two-phase configurations at coexistence. We then find configurations \(\mathbf{Q}(\mathbf{r})\) that are not uniform, and that minimize Eq. \eqref{FreeE} subject to the constraint \eqref{MicroQ}.

\begin{figure}
	\includegraphics[width = \columnwidth]{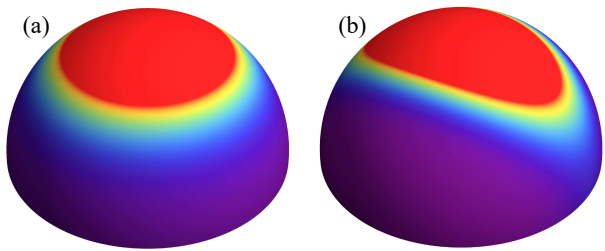}
	\caption{Examples of the probability distribution, \(p(\bm{\xi})\) of Eq. \eqref{prob}, on the sphere spanned by \(\bm{\xi}\) for (a) a uniaxial configuration and (b) a biaxial configuration. Note that the probability distribution involves a uniaxial molecule, but a biaxial order parameter can occur for a probability distribution with biaxial second moment. Only northern hemispheres are displayed since the probability distribution is symmetric about the equator due to the symmetry of the molecules. For these plots, (a) \(\bm{\Lambda} = 4 \diag(-1,\,-1,\,0.5)\) and (b) \(\bm{\Lambda} = 10\diag(-0.25,\,-1,\,0.25)\).}
	\label{fig:probdist}
\end{figure}

The entropy, Eq. \eqref{deltaS}, can be maximized, subject to the constraint \eqref{MicroQ}, by introducing a tensor of Lagrange multipliers, \(\bm{\Lambda}(\mathbf{r})\), for each component of the constraint \cite{ball10,katriel86}. The resulting probability that maximizes the entropy is given by
\begin{align} 
p(\bm{\xi};\mathbf{r}) &= \frac{\exp[\bm{\xi}^T \bm{\Lambda}(\mathbf{r})\bm{\xi}]}{Z[\bm{\Lambda}(\mathbf{r})]} \label{prob} \\
Z[\bm{\Lambda}(\mathbf{r})] &= \int_{S^2} \exp[\bm{\xi}^T \bm{\Lambda}(\mathbf{r})\bm{\xi}] \, d\bm{\xi} \label{Z}
\end{align}
where \(Z\) can be interpreted as a single particle partition function. Fig. \ref{fig:probdist} shows graphical examples of the probability distribution on the unit sphere. We mention that the single particle partition function can only be computed numerically, and hence the minimization procedure described next has to be carried out numerically in its entirety.

The minimization of \(F\) in Eq. \eqref{FreeE} with \(p(\bm{\xi};\mathbf{r})\) given by Eqs. \eqref{prob} and \eqref{Z} is therefore reformulated in terms of two tensor fields on the domain, \(\mathbf{Q}(\mathbf{r})\) and \(\bm{\Lambda}(\mathbf{r})\) (from here on the dependence on \(\mathbf{r}\) will be dropped for brevity). \(\bm{\Lambda}\) acts as an effective interaction field which mediates interactions among molecules. Substituting Eq. \eqref{prob} into the constraint, Eq. \eqref{MicroQ}, leads to a relation between \(\mathbf{Q}\) and \(\bm{\Lambda}\):
\begin{equation} \label{consist}
\mathbf{Q} + \frac{1}{3} \mathbf{I} = \frac{\partial \ln Z[\bm{\Lambda}]}{\partial \bm{\Lambda}}.
\end{equation}
It has been shown that if the eigenvalues of \(\mathbf{Q}\) approach the endpoints of their physically admissible values, both \(\bm{\Lambda}\) and the free energy diverge. This feature is not present in the LdG theory, which can lead to nonphysical configurations for certain choices of the elastic energy, \(f_e\), in Eq. \eqref{H} \cite{ball10,bauman16}.

The fields \(\mathbf{Q}\) and \(\bm{\Lambda}\) that minimize Eq. \eqref{FreeE} and satisfy Eq. \eqref{consist} are the equilibrium configuration for a given set of boundary conditions. In the next section we describe a computational implementation of the model presented here.

\section{\label{compMeth}Computational Method}
\subsection{Molecular Theory}
To find the configuration \(\mathbf{Q}\) that minimizes the free energy of the molecular field theory we numerically solve the differential equations \(\delta F / \delta \mathbf{Q} = 0\). This, in principle, is a system of nine equations. However, since \(\mathbf{Q}\) is traceless and symmetric, there are only five degrees of freedom. The eigenvalues of \(\mathbf{Q}\) describe two degrees of freedom since \(\mathbf{Q}\) is traceless. The eigenvectors of \(\mathbf{Q}\) form an orthonormal frame (since \(\mathbf{Q}\) is symmetric) which accounts for the other three degrees of freedom: the first vector has two degrees of freedom since it is a unit vector, the second vector has one degree of freedom since it is a unit vector and must be orthogonal to the first vector, and the third vector is determined from the other two vectors since it must be orthogonal to both. The eigenvalues are related to the amount of order in the system, while the eigenvector which corresponds to the largest eigenvalue is the director, \(\mathbf{n}\). This is illustrated in Fig. \ref{fig:probdist} which shows the probability distribution for molecules with a director along the z-axis. Fig. \ref{fig:probdist}a shows a uniaxial configuration in which two of the eigenvalues are degenerate, leading to arbitrary eigenvectors in the xy-plane. It is possible for the probability distribution to be of the form in Fig. \ref{fig:probdist}b in which the director is still along the z-axis, but all three eigenvalues are distinct. In this case, we call the probability distribution biaxial since it leads to a second moment, \(\mathbf{Q}\), that is biaxial. It is known that biaxiality of the order parameter is important near defects and at interfaces in systems of uniaxial molecules as modeled by the LdG free energy \cite{pismen99,popanita97,mottram14}. Despite the uniaxial character of the molecules, Eq. (\ref{MicroQ}), the molecular theory detailed here can accommodate biaxial order.

Local biaxial order will be parametrized as
\begin{equation} \label{Qdef}
\mathbf{Q} = S(\mathbf{n} \otimes \mathbf{n} - \frac{1}{3} \mathbf{I}) + P(\mathbf{m} \otimes \mathbf{m} - \bm{\ell} \otimes \bm{\ell})
\end{equation}
where \(\{\mathbf{n},\mathbf{m},\bm{\ell}\}\) are an orthonormal triad of vectors. This representation explicitly includes the five degrees of freedom of \(\mathbf{Q}\), namely, three for the orthonormal set of vectors and two for the amplitudes \(S\) and \(P\). In addition to \(\mathbf{n}\) being the director, \(S\) represents the amount of uniaxial order, and \(P\) the amount of biaxial order. That is, \(S = (3/2)\, q_1\) and \(|P| = (1/2)\, (q_2 - q_3)\) where \(q_i\) are the eigenvalues of \(\mathbf{Q}\), and \(q_3 \leq q_2 \leq q_1\).

Because we are primarily concerned with experiments in thin nematic films, we further reduce the degrees of freedom of \(\mathbf{Q}\) by only considering spatial variation in at most two dimensions. If we write \(\mathbf{n} = (\cos \phi, \,\sin \phi, \,0)\), \(\mathbf{m} = (-\sin \phi, \,\cos\phi,\,0)\), and \(\bm{\ell} = (0,\,0,\,1)\), where \(\phi\) is the angle the director makes with the x-axis, we need only one degree of freedom to describe the eigenframe of \(\mathbf{Q}\). We can then further simplify the computations by transforming to the auxiliary variables \cite{sen86}
\begin{align}
\eta &= S - \frac{3}{2}(S - P) \sin^2 \phi \nonumber \\
\mu &= P + \frac{1}{2}(S - P) \sin^2 \phi \label{aux} \\
\nu &= \frac{1}{2}(S - P) \sin 2\phi. \nonumber
\end{align}
This transformation is equivalent to expressing \(\mathbf{Q}\) in terms of a new basis for traceless, symmetric matrices. While we do this for ease of computation, we can transform back to the original parametrization after calculating the eigenvalues and eigenvectors of \(\mathbf{Q}\). Although all of our calculations are conducted with the set \(\{\eta,\mu,\nu\}\), we will present our results in terms of the more physically intuitive \(S\), \(P\), and \(\phi\). 

The tensor order parameter in this representation is
\begin{equation} \label{auxQ}
\mathbf{Q} = 
\begin{bmatrix}
\frac{2}{3}\eta & \nu & 0\\
\nu & -\frac{1}{3} \eta + \mu & 0 \\
0 & 0 & -\frac{1}{3} \eta - \mu
\end{bmatrix}.
\end{equation}
We can now substitute Eq. \eqref{auxQ} into Eq. \eqref{MicroQ} to write the constraint in terms of \(\eta\), \(\mu\), and \(\nu\). Following the procedure of Section \ref{Model}, we introduce three Lagrange multipliers \(\Lambda_1\), \(\Lambda_2\), and \(\Lambda_3\) corresponding to \(\eta\), \(\mu\) and \(\nu\) respectively, and a partition function
\begin{equation} \label{auxZ}
Z[\Lambda_1,\Lambda_2,\Lambda_3] = \int_{S^2} \exp\bigg[\frac{3}{2} \Lambda_1 \xi_1^2 + \Lambda_2(\frac{1}{2} \xi_1^2 + \xi_2^2) + \Lambda_3\xi_1 \xi_2\bigg]\, d\bm{\xi}
\end{equation}
while the relation from Eq. \eqref{consist} manifests itself as the three equations
\begin{align}
\frac{\partial \ln Z}{\partial \Lambda_1} &= \eta + \frac{1}{2} \nonumber \\
\frac{\partial \ln Z}{\partial \Lambda_2} &= \mu + \frac{1}{2} \label{auxConstraint} \\
\frac{\partial \ln Z}{\partial \Lambda_3} &= \nu \nonumber
\end{align}
that implicitly relate the variables \(\eta\), \(\mu\), and \(\nu\) to the Lagrange multipliers. Note that since \(Z[\Lambda_1,\Lambda_2,\Lambda_3]\) cannot be obtained analytically, relation \eqref{auxConstraint} can only be solved numerically. The free energy, Eq. \eqref{FreeE}, is rewritten as
\begin{widetext}
\begin{equation} 
F = \int_{\Omega} \Big( f_b(\eta,\mu,\nu,\Lambda_1,\Lambda_2,\Lambda_3) + f_e(\eta,\mu,\nu,\nabla \eta,\nabla \mu,\nabla\nu)\Big) \, d\mathbf{r}
\end{equation}
where \(f_b\) is a bulk free energy density that does not depend on gradients of the fields. Written explicitly,
\begin{equation} \label{FreeEAux}
f_b = -2\alpha \big(\frac{1}{3}\eta^2 + \mu^2 + \nu^2\big) + \\
n k_B T \Big(\Lambda_1 \big(\eta + \frac{1}{2}\big) + \Lambda_2 \big(\mu + \frac{1}{2}\big) + \Lambda_3 \nu + \ln(4\pi) - \ln Z[\Lambda_1,\Lambda_2,\Lambda_3]\Big).
\end{equation}
\end{widetext}

We will focus in this paper on an isotropic elastic energy \(f_e = L \partial_k Q_{ij} \partial_k Q_{ij}\) where repeated indices are summed, and \(L\) is the elastic constant. This is the `one constant approximation' so that mapping this elastic energy to the Oseen-Frank elastic energy yields the same value for all three elastic constants \cite{longa87}. Written in terms of the auxiliary variables we have
\begin{equation} \label{elasticE}
 f_e = 2 L \big(\frac{1}{3} |\nabla \eta|^2 + |\nabla \mu|^2 + |\nabla \nu|^2\big).
\end{equation}
Before deriving the differential equations to be solved we redefine quantities in a dimensionless way:
\begin{equation} \label{units}
\tilde{f_b} = \frac{f_b}{n k_B T}, \quad \tilde{f_e} = \frac{f_e}{n k_B T}, \quad \tilde{x} = \frac{x}{\xi_{MS}}, \quad \tilde{L} = \frac{L}{\xi_{MS}^2 n k_B T}
\end{equation}
where \(\xi_{MS}\) is a length scale which we set by defining the value of the dimensionless parameter \(\tilde{L}\) instead. For the rest of the paper the tildes are omitted for brevity.

To derive the equilibrium equations, we note that Eq. \eqref{auxConstraint} relates \(\eta\), \(\mu\), and \(\nu\) as functions of \( \{ \Lambda_i \} \) through the unknown single particle partition function. It has been shown that these relations are invertible when \(\eta\), \(\mu\), and \(\nu\) give physical eigenvalues of \(\mathbf{Q}\) \cite{katriel86}. We can then regard \(\Lambda_1\), \(\Lambda_2\), and \(\Lambda_3\) as functions of \(\eta\), \(\mu\), and \(\nu\) via the inverse of Eq. \eqref{auxConstraint}. Although an analytic inverse does not exist we can numerically invert this equation using a Newton-Raphson method. We create a MATLAB scattered interpolant from values given by the Newton-Raphson method. We select interpolant points from the values \(0 \leq S \leq 0.7\), \(0 \leq P \leq 0.1\), and \(-\pi/2 \leq \phi \leq \pi/2\) with \(\Delta S = \Delta P = 0.05\) and \(\Delta \phi = 0.0245\). These values are then transformed to \(\eta\), \(\mu\), and \(\nu\) through Eqs. \eqref{aux} and the Newton-Raphson method is run using these values to find \(\Lambda_i\) for the chosen interpolant points. The MATLAB scattered interpolant is then created and used in the numerical minimization procedure. The Euler-Lagrange equations are derived by taking the variations of Eqs. \eqref{FreeEAux} and \eqref{elasticE} with respect to \(\eta\), \(\mu\), and \(\nu\) while using Eqs. \eqref{auxConstraint} to simplify. The dimensionless equations are
\begin{align}
\frac{4}{3} L \nabla^2 \eta = \Lambda_1 - \frac{4}{3}\frac{\alpha}{n k_B T} \eta \nonumber \\
4 L \nabla^2 \mu = \Lambda_2 - 4 \frac{\alpha}{n k_B T} \mu \label{ELEqs} \\
4 L \nabla^2 \nu = \Lambda_3 - 4 \frac{\alpha}{n k_B T} \nu \nonumber
\end{align}
where, again, \(\Lambda_i\) are numerically calculated as functions of \(\eta\), \(\mu\), and \(\nu\). Eqs. \eqref{ELEqs} are the central equations of this study and are solved numerically in the following section for various cases of interest.

To numerically solve Eqs. \eqref{ELEqs} we use a finite differencing scheme. For one-dimensional configurations, an implicit backward Euler method is used with 129 discrete points and time step \(\Delta t = 0.1 \Delta x^2\). For two-dimensional configurations a Gauss-Seidel relaxation method with \(257^2\) discrete points is used \cite{press02}. We iterate until the calculated energy of a configuration fails to change to within \(10^{-7}\). We check that the calculated energy of the initial condition is larger than the energy of the final configuration. In all cases we use Dirichlet boundary conditions that depend on the case being studied, as described in the relevant section. The MATLAB code used for the numerical solutions can be found in Ref. \cite{schimming20}.

\subsection{Landau-de Gennes Theory}

Here, we summarize the conventions and notation used in the calculations to compare the LdG free energy with the molecular field theory presented in the previous section. The bulk energy density is of the form
\begin{equation} \label{LdGE}
f_{LdG} = \frac{1}{2} a(T - T^*) \Tr [\mathbf{Q}^2] - \frac{1}{3} B \Tr [\mathbf{Q}^3] + \frac{1}{4} C \big( \Tr [\mathbf{Q}^2] \big)^2
\end{equation}
where \(a\), \(B\), and \(C\) are material parameters, and \(T^*\) is the temperature at which the isotropic phase loses its stability. We use the same elastic free energy defined above when comparing to the molecular field theory as well. For the sake of computation, we define the following dimensionless quantities:
\begin{equation} \label{LdGunits}
\tilde{f}_{LdG} = \frac{f_{LdG}}{C}, \quad \tilde{f_e} = \frac{f_{LdG}}{C}, \quad \tilde{x} = \frac{x}{\xi_{LdG}}, \quad \tilde{L} = \frac{L}{\xi_{LdG}^2 C}
\end{equation}
which leaves \(a (T - T^*)/C\), \(B/C\), and \(\tilde{L}\) as dimensionless parameters for the model. \(\xi_{LdG}\) here is a length scale for the model defined by the value of \(\tilde{L}\) similar to \(\xi_{MS}\) in Eq. \eqref{units}. As before, the tilde is subsequently dropped for brevity.

Computations are done using the same auxiliary variables defined in Eq. \eqref{aux} with the same finite difference scheme outlined above to solve the Euler-Lagrange equations resulting from \(f_{LdG}\).

\section{\label{results}Results}

\begin{figure}
	\includegraphics[width = \columnwidth]{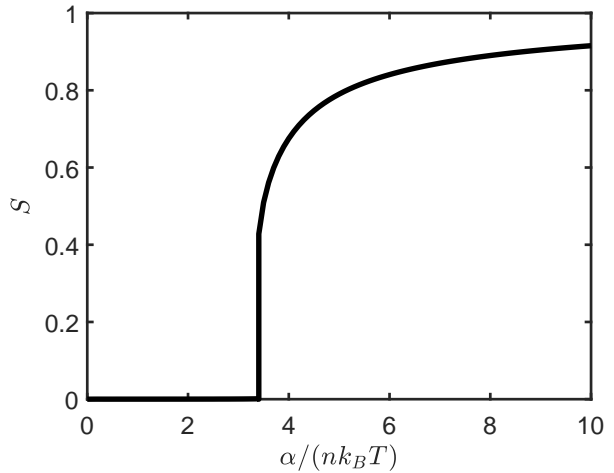}
	\caption{Equilibrium value of the uniaxial order, \(S\), versus the parameter \(\alpha / (n k_B T)\). At high \(T\), the system is in an isotropic phase, while at low \(T\) the system is in a uniaxial nematic phase. A first order phase transition occurs at \(\alpha / (n k_B T) \approx 3.4049\).}
	\label{fig:phaseD}
\end{figure}

\subsection{Uniform Configuration and Bulk Free Energy}
We first check our numerical method and methodology with known results for the Maier-Saupe free energy. As mentioned above, this model should be equivalent to the Maier-Saupe model in the case of a uniform system, \(f_e = 0\). In this case, it has been shown that minimizers of the bulk free energy, Eq. \eqref{FreeEAux}, will be uniaxial states \cite{ball10}. Thus, because we are considering a uniform system, the choice of director is arbitrary. We choose \(\phi = 0\) for this analysis so the auxiliary variables defined by Eq. \eqref{aux} give \(\eta = S\), \(\mu = P\), and \(\nu = 0\). Further, since we know the system will be uniaxial we can take \(\mu = P = 0\). One can show that this implies \(\Lambda_2 = \Lambda_3 = 0\) from Eq. \eqref{auxConstraint}. 

Because the system is uniform, \(S\) is constant, and hence \(\nabla^2 S = 0\). Defining \(S_N\) as the value of \(S\) in uniform equilibrium, we find, from Eq. \eqref{ELEqs}:
\begin{equation} \label{equilLam}
\Lambda_1 = \frac{4}{3} \frac{\alpha}{n k_B T} S_N
\end{equation}
which is a well known result for the Maier-Saupe model when \(\Lambda_1\) is regarded as an effective interaction strength \cite{deGennes75,maier59,selinger16}. We then substitute Eq. \eqref{equilLam} into Eq. \eqref{FreeEAux} and numerically minimize it to find the value of \(S\) in equilibrium for a uniform system. Fig. \ref{fig:phaseD} shows \(S_N\) as a function of \(\alpha / (n k_B T)\). At high temperatures, the equilibrium phase is isotropic with \(S=0\). At low temperatures a uniaxial nematic phase is stable with \(S = S_N\). A first order phase transition occurs at \(\alpha / (n k_B T) \approx 3.4049 \) with \(S_N = 0.4281\). The diagram of Fig. \ref{fig:phaseD} agrees with previous studies of the Maier-Saupe model which has been used successfully to describe phase transitions in experiments \cite{selinger16}.

\begin{figure}
	\includegraphics[width = \columnwidth]{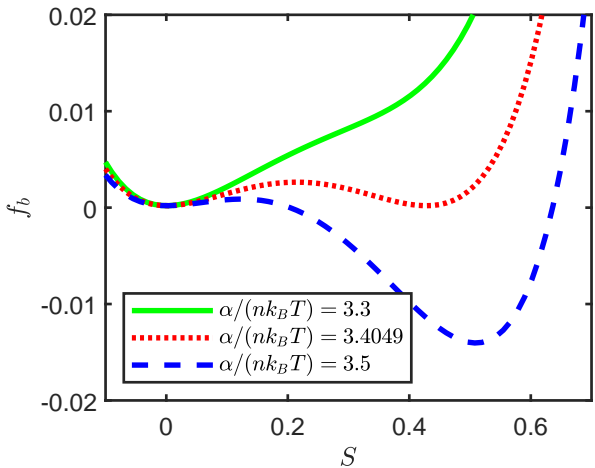}
	\caption{Bulk free energy density as a function of the uniaxial order, \(S\) for three values of the parameter \(\alpha / (n k_B T)\). As \(\alpha / (n k_B T)\) increases, the free energy becomes non-convex, leading to coexistence between the isotropic and nematic phases.}
	\label{fig:bulkE}
\end{figure}

We can further elucidate the nature of the molecular field theory by examining the bulk free energy density, Eq. \eqref{FreeEAux}, restricted to a uniaxial configuration. For a uniform, uniaxial system, the free energy density is
\begin{equation} \label{bulkDens}
f_b(S) = -\frac{2}{3}\frac{\alpha}{n k_B T} S^2 + \Lambda_1 \Big( S + \frac{1}{2} \Big) - \ln Z[\Lambda_1] + \ln(4 \pi)
\end{equation}
where \(\Lambda_1\) is calculated as a function of \(S\) through Eq. \eqref{auxConstraint}. This function is plotted in Fig. \ref{fig:bulkE} for three different values of \(\alpha / (n k_B T)\). As \(\alpha / (n k_B T)\) increases we find that \(f_b\) becomes non-convex, leading to a coexistence region in the phase diagram, and a first order phase transition. It is well known that these features are also present in the LdG free energy of Eq. \eqref{LdGE} \cite{gramsbergen86}. The primary difference between LdG and the Maier-Saupe theory is that in the latter \(f_b\) diverges when \(S = -1/2\) or \(S = 1\), that is, when the eigenvalues leave the physical range. The non-convexity obtained agrees with similar plots for the Maier-Saupe free energy in Ref. \cite{selinger16}.

The non-convexity and similarity of the bulk free energy to LdG suggest that there should exist stable interfacial configurations at coexistence as well as stable solutions for topological defects in the nematic phase. In the following three subsections we demonstrate just this and compare to results given by LdG theory. 

\subsection{Planar Isotropic-Nematic Interface}
We consider a one-dimensional configuration with a planar interface in which the order parameter \(\mathbf{Q}(\mathbf{r}) = \mathbf{Q}(x)\). We solve Eqs. \eqref{ELEqs} on a domain of size \(\mathcal{L} = 100 \xi_{MS}\) with Dirichlet boundary conditions where \(S = S_N\) at \(x = -50\xi_{MS}\) and \(S = 0\) at \(x = 50\xi_{MS}\). We set \(\alpha / (n k_B T) = 3.4049\) and \(S_N = 0.4281\) so that the isotropic and nematic bulk phases coexist. An important note is that since we are using the \lq\lq one-constant approximation'' for the elastic free energy there are no anisotropic effects, such as anchoring, in our analysis. It is known that anisotropy changes the width of an interface for different director orientations, however, because we are only considering isotropic terms here the structure of the interfacial profile should not change if the angle of the director in the nematic phase, \(\phi\), is changed \cite{popanita97}. 

\begin{figure}
	\includegraphics[width = \columnwidth]{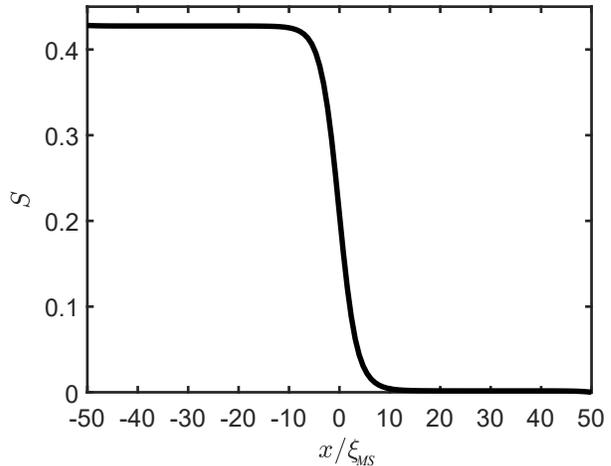}
	\caption{\(S\) as a function of position for a one-dimensional interface. Dirichlet boundary conditions maintain \(S = S_N\) at the left boundary while \(S = 0\) at the right boundary. \(L = 1\) for this configuration.}
	\label{fig:interface} 
\end{figure}

Fig. \ref{fig:interface} shows the equilibrium uniaxial order parameter \(S\) for \(\phi=0\). We find a smooth, diffuse interface with \(P=0\), that is, no biaxiality. We also find that changing the angle of the director does not change the solution, as expected. We can calculate the width of the interface by finding the points where \(S = 0.1 S_N\) and \(S = 0.9 S_N\) and define them as \(x_1\) and \(x_2\) respectively. Then we define the width as \(x_1 - x_2\).  

In order to compare with the LdG free energy, Eq. \eqref{LdGE}, we recall that the interfacial profile for this configuration is known exactly
\begin{equation} \label{SLdG}
S_{LdG}(x) = \frac{S_N}{2} \bigg(1 - \tanh \Big(\frac{x}{w_{LdG}}\Big)\bigg)
\end{equation}
with
\begin{equation} \label{w}
w_{LdG} = \frac{6\sqrt{6}}{B / C}\sqrt{L}
\end{equation} 
which sets the width of the interface. This implies that \((x_1 - x_2) \propto \sqrt{L}\). One can similiarly show that the bulk energy contribution, i.e. the bulk contribution to the surface tension, \(\sigma \propto \sqrt{L}\).

\begin{figure}
	\includegraphics[width = \columnwidth]{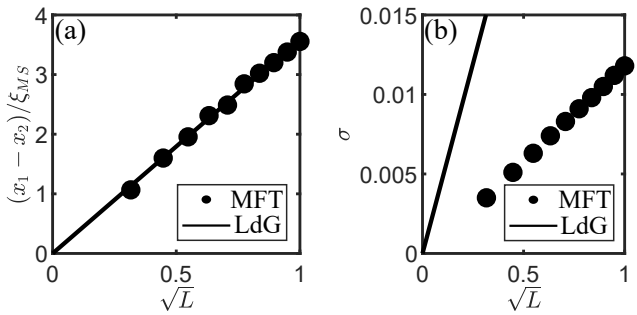}
	\caption{(a) Interface width and (b) bulk surface tension versus \(\sqrt{L}\). Dots represent the molecular field theory (MFT) computations while the solid lines are derived from the analytical solution for LdG, Eq. \eqref{SLdG}, with \(B/C = 9\). Both the interface width and excess free energy (i.e. surface tension) scale linearly with the parameter \(\sqrt{L}\), the same scaling relationship as that of Landau-de Gennes.}
	\label{fig:widths}
\end{figure}

With this in mind, we compare the scaling of the molecular field theory solutions that we obtain with \(\sqrt{L}\). To this end, we find the interface widths and bulk surface tensions for solutions to Eqs. \eqref{ELEqs} for a variety of values of \(L\). The bulk surface tension is found by numerically integrating the bulk free energy density, Eq. \eqref{FreeEAux}. Interface widths and bulk surface tensions are plotted in Fig. \ref{fig:widths} for both the molecular field theory and LdG. We find both \((x_1 - x_2) \propto \sqrt{L}\) and \(\sigma \propto \sqrt{L}\) for the molecular field theory. Note that the LdG solution allows additional tuning via the parameter \(B/C\), which we have set to 9 in Fig. \ref{fig:widths}. In Fig. \ref{fig:widths}b the discrepency between the LdG solution and the molecular field theory computations highlights that even if the widths of LdG interfaces are tuned to be similar to those of the molecular field theory, the surface tensions cannot be, and vice versa.

We note that the similarity in bulk free energy landscape likely leads to the similarity in solutions for LdG and the molecular field theory. Anisotropic effects have yet to be analyzed for our model, for which it is known for LdG there is nonzero biaxiality at interfaces \cite{popanita97}. This will be the subject of a future study.

\begin{figure*}
	\includegraphics[width = \textwidth]{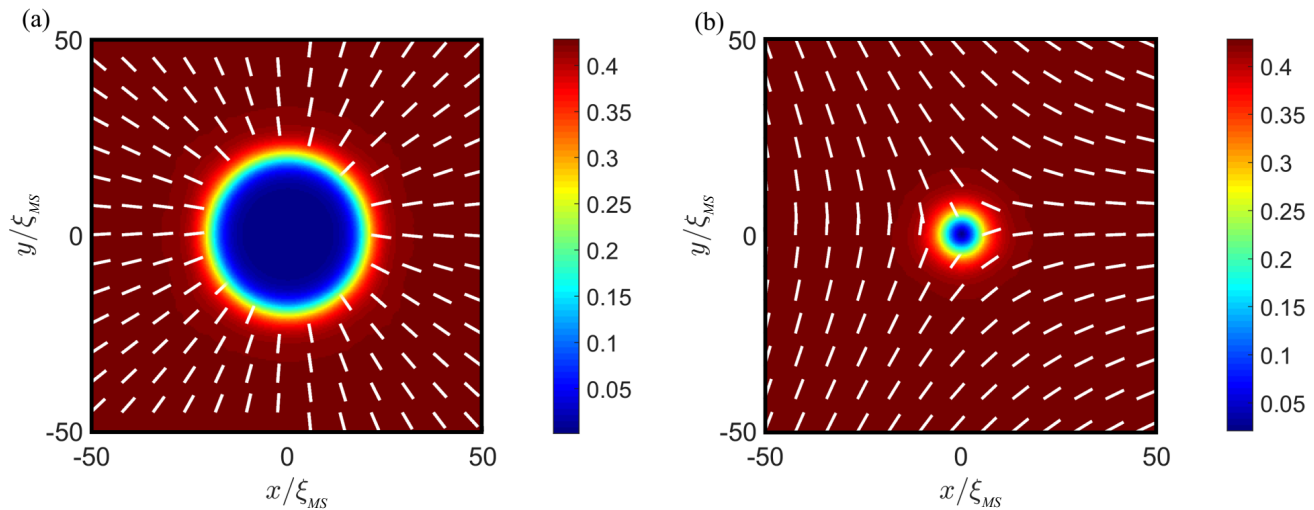}
	\caption{Plots of \(S(x,y)\) for (a) a tactoid with \(m=1\) director configuration at the outer boundary and (b) tactoid with \(m=-1 / 2\) director configuration at the outer boundary. The radius in (a) is \(R/\xi_{MS} = 19.92 \pm 0.2\) and the radius in (b) is \(R/\xi_{MS} = 4.59 \pm 0.2\). The smaller size of the \(m = -1/2\) tactoid is due to the director distortion energy's \(m^2\) dependence. For both computations \(L=1\).}
	\label{fig:tactoids}
\end{figure*}

\subsection{Tactoids}
We consider a two-dimensional square domain of size \(\mathcal{L} = 100 \xi_{MS}\). We set \(S=S_N\), \(P=0\), and \(\phi = m \theta\) at the outer boundary, where \(\theta\) is the polar angle and \(m\) is the winding number of \(\phi\). We set \(\alpha / (n k_B T) = 3.4049\) and \(L = 1\). As initial conditions we set \(S = 0\) within a disc centered at the origin of radius \(R = 15 \xi_{MS}\).

By \lq\lq tactoid" we refer to a two-phase domain separated by an interface. In the isotropic region \(S = P = 0\). We consider distorted boundary conditions to ensure an interface forms in the simulation. Because the director can vary as a function of position in two dimensions, the boundary conditions imposed will change the size and shape of the object under consideration. Since we are only considering isotropic gradients in the elastic free energy, there is no anchoring term at the interface, i.e. there is not a difference in energy based on the orientation of the molecules relative to the interface. Thus, we expect the tactoids to be cylindrical. The topology of the boundary conditions does impact the size of the tactoids, however. This is due to a balance between two energies: the surface tension, which in two dimensions is proportional to \(R\), the radius of the tactoid, and the elastic energy in the nematic region from Oseen-Frank which is proportional to \(m^2 \ln(\mathcal{L}/R)\). Due to the symmetry of the molecules, half integer \(m\) is allowed and costs four times less director distortion energy than integer \(m\). Hence, we expect that tactoids with integer boundary conditions should be approximately four times larger than those with half integer boundary conditions.

In Fig. \ref{fig:tactoids}, we show equilibrium configurations for boundary conditions with \(m=1\) and \(m=-1/2\). In both cases an isotropic region with \(S = P = 0\) is present at the center of the computational domain. As expected, both configurations are cylindrical in shape and we find that \(R/ \xi_{MS} =19.92 \pm 0.2\) for the \(m=1\) configuration and \(R / \xi_{MS} = 4.59 \pm 0.2\) for the \(m=-1/2\) configuration. To find the radii we take a cut from the center of the tactoid to the outer boundary and find the point where \(S = 0.5 S_N\). It should be noted that LdG, in the one-constant approximation in elastic energy, gives similar results in terms of the size and shape of tactoids. 

It is known for the LdG bulk free energy with anisotropic elastic free energies that the shape of the tactoids also changes due to anchoring at the interface \cite{golovaty19}. Anisotropic effects on the shape of tactoids in the molecular field theory will be the subject of a future study. 

\subsection{Nematic Disclinations}
We consider next the case of disclination lines in thin films. We consider a two-dimensional square of size \(\mathcal{L} = 10 \xi_{MS}\). For all calculations \(L=1\) and \(\alpha / (n k_B T) > 3.4049\), so nematic ordering is energetically advantageous. At the outer boundary we fix the system to be uniaxial (\(P=0\)) and fix the director orientation, \(\phi = (-1/2) \theta\). The initial configuration is \(S(r) = S_N\big(1 - \exp(r / 2)\big)\) with \(P=0\) everywhere.

\begin{figure*}
	\includegraphics[width = \textwidth]{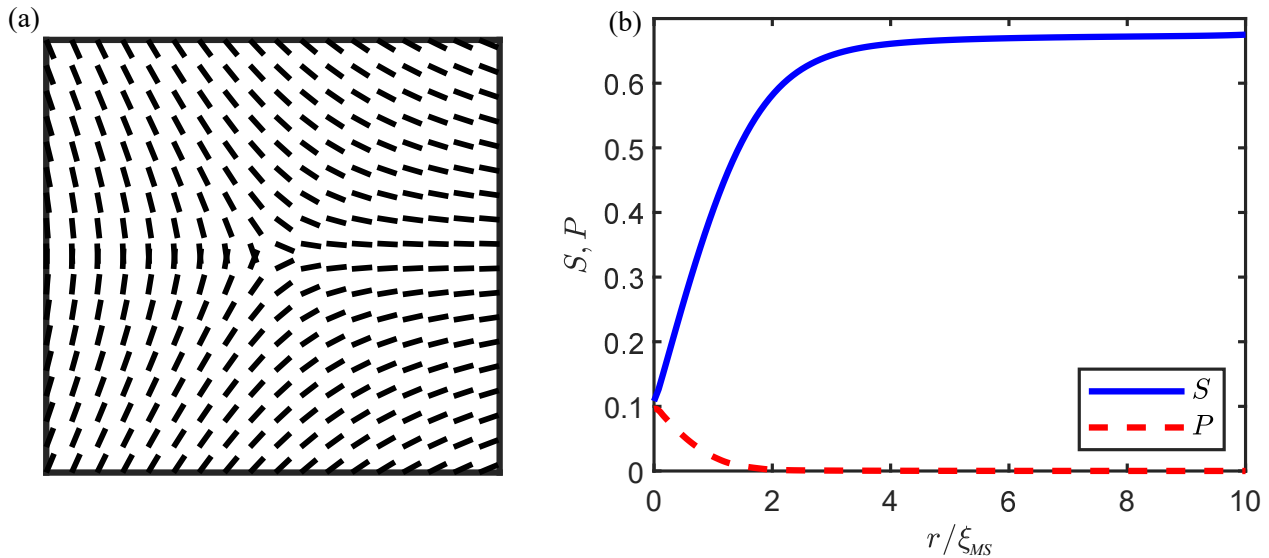}
	\caption{(a) Director profile and (b) radial plots of the uniaxial order \(S\) and the biaxial order \(P\) for a nematic disclination. The spatial extent of biaxiality is on the order of the radius of the disclination core. Here, \(\alpha / (n k_B T) = 4\) and \(L= 1\).}
	\label{fig:radDefect}
\end{figure*}

In Fig. \ref{fig:radDefect} we show the director profile, and the radial profile of equilibrium \(S\) and \(P\) from the center of a disclination to the boundary of the domain for the parameter \(\alpha / (n k_B T) = 4\). For the director, \(\phi = -(1/2) \theta\) outside the core. Much like solutions for the LdG free energy, we see a disclination core that is biaxial \cite{meiboom82,schopohl87}. The biaxiality of the core was explained topologically by Lyuksyutov, assuming a LdG bulk free energy \cite{lyuksyutov78}. Using this free energy for analysis, one can define a \lq\lq biaxial length'' scale for the disclinations, \(R_b \approx \sqrt{K/(B S^3)}\), where \(K\) is on the order of the Frank constants and \(B\) is the parameter associated with the cubic term in the LdG bulk energy, Eq. \eqref{LdGE}. For distances from the core smaller than \(R_b\), the elastic energy becomes comparable to the cubic term in the LdG free energy and the system can remove the elastic singularity by becoming biaxial, since a biaxial order parameter can remove the singularity. We note that at the core, \(S = P\) in both models. Using the parametrization from Eq. \eqref{Qdef}, one can show that this is interpreted as a uniaxial order parameter, but for a disc if \(S > 0\) or a rod aligned with the z-axis if \(S < 0\). For both models, \(S > 0\) at the core. Thus, we interpret the biaxial solution as a macroscopic \lq\lq transformation'' of rods far away from the core to discs at the core. Microscopically, the probability distribution describing individual molecules becomes more and more spread out in the x-y plane in an attempt to alleviate the elastic energy singularity.

We emphasize that it is not obvious that the molecular field theory should give biaxial core solutions for the disclinations since, by construction, the model is markedly different from LdG. While LdG is an expansion of a macroscopic order parameter, the model here is based on a microscopic description. Because of this, it is difficult to quantitatively compare the solutions for the disclinations given by the two models. While we note that the spatial extent of the biaxiality for the disclinations is on the order of the radius of the defects, there is not a cubic term in the free energy to define a length such as \(R_b\). Instead, this behavior is induced by the single particle partition function which appears in Eq. \eqref{FreeEAux} since the Maier-Saupe energy is purely quadratic in \(\mathbf{Q}\).

\begin{figure}
	\includegraphics[width = \columnwidth]{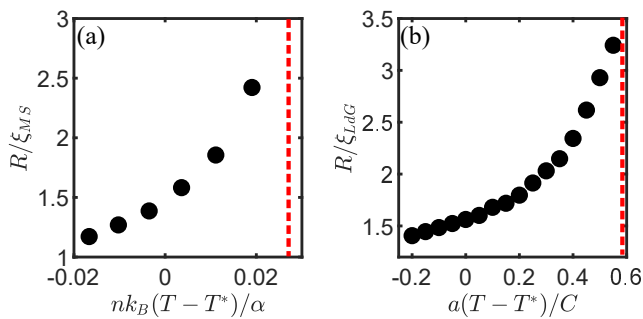}
	\caption{Radius of disclinations plotted as a function of temperature for (a) the molecular field theory of section \ref{Model} and (b) the Landau-de Gennes model. \(T^*\) is the temperature where the isotropic phase loses its metastability, while the dotted line on the plots indicate where coexistence between phases is for the respective model. For the molecular field theory we use \(L = 1\), and for Landau-de Gennes \(L = 1\) and \(B / C = 4\) for all simulations.}
	\label{fig:defectSize}
\end{figure}

Another aspect of the disclinations that we can compare, at least qualitatively, to the LdG model is the scaling of the radius of disclinations with temperature. To find the radius, we take a cut from the center of the disclination to the boundary and find the point where \(S - P = S_N ( 1 - e^{-1})\). The results are plotted in Fig. \ref{fig:defectSize}. We show both the scaling for the molecular field theory and for results given by LdG. It can be seen that the scaling is similar for both models in a wide range of temperatures up to the coexistence temperature, where the isotropic phase becomes energetically favorable.

We are currently investigating the effects of anisotropic elastic free energies on disclinations. It is known that the director structure becomes less symmetric away from the disclination core if the Frank constants for bend and splay are not equal, and recent experiments have found anisotropic core structures \cite{zhou17}.

\section{\label{Conclusion}Conclusion}
In this work, we have presented a computational implementation of the model of reference \cite{ball10}. We show that the model can be interpreted as replacing direct interactions between molecules via an effective interaction field \(\bm{\Lambda}\) in the mean field approximation. Further, we investigate the similarity between the free energy of this molecular field theory and the LdG free energy and compare solutions given by both for the cases of interfaces, tactoids, and topological defects. We find that all have qualitatively similar results which is an interesting result given that the construction of the two models is very different.

This model allows for a more fundamental understanding of the underlying microscopic and mesoscopic physics at play, and can serve as an alternative to the LdG free energy when describing systems with inhomogeneous ordering. The extension of the Maier-Saupe model to a field theory allows us to understand not just the phase transition but also inhomogeneous configurations, and can possibly be used to describe experiments like those of Refs. \cite{zhou17,kim13}. 

Moving forward, we are currently investigating the results of adding anisotropy to the elastic free energy, which has been done to some extent for the LdG model \cite{golovaty19}. Importantly, however, one can consider in this framework the values of the elastic constants for chromonics that have been determined experimentally \cite{zhou14}, while avoiding boundedness issues in LdG theory when bend and splay constants are different. Further, because of the microscopic nature of the model, one can, in principle, use a more physically realistic Hamiltonian to describe the molecular system, as opposed to the effective Maier-Saupe Hamiltonian that is used here. One can also generalize the computations to more complex molecules, such as plate-like molecules, by modifying Eq. \eqref{MicroQ}.

\begin{acknowledgments}
We are indebted to Shawn Walker and Sergij Shiyanovskii for useful discussions. This research is supported by the National Science Foundation under contract DMR-1838977, and by the Minnesota Supercomputing Institute.
\end{acknowledgments}

%

\end{document}